\documentclass[twocolumn,groupedaddress,superscriptaddress]{revtex4-1}
\usepackage{eurosym}
\usepackage{amsfonts}
\usepackage{amsmath}
\usepackage{amssymb}
\usepackage{graphicx}
\usepackage{float}

\begin{document}

% La$_{1.875}$Sr$_{0.125}$CuO$_4$

\title{The nature of the phase transition in the cuprates as revealed by a magnetic field free stiffness meter}

\author{Itzik Kapon}
\email[E-mail: ]{itzikapon@gmail.com}
\affiliation{Department of Physics, Technion - Israel Institute of Technology, Haifa, 3200003, Israel}

\author{Zaher Salman}
\affiliation{Laboratory for Muon Spin Spectroscopy, Paul Scherrer Institute, CH 5232 Villigen PSI, Switzerland}

\author{Thomas Prokscha}
\affiliation{Laboratory for Muon Spin Spectroscopy, Paul Scherrer Institute, CH 5232 Villigen PSI, Switzerland}

\author{Nir Gavish}
\affiliation{Department of Mathematics, Technion - Israel Institute of Technology, Haifa, 3200003, Israel}

\author{Amit Keren}
\email[E-mail: ]{keren@physics.technion.ac.il}
\affiliation{Department of Physics, Technion - Israel Institute of Technology, Haifa, 3200003, Israel}

\date{\today }
%\linenumbers

\begin{abstract}

A new method to measure the superconducting stiffness tensor $\overline{\rho}_s$, without subjecting the sample to magnetic field, is applied to La$_{1.875}$Sr$_{0.125}$CuO$_4$ (LSCO). The method is based on the London equation $\bf{J}=-\overline{\rho}_s \bf{A}$, where $\bf{J}$ is the current density and $\bf{A}$ is the vector potential. Using rotor free $\bf{A}$ and measuring $\bf{J}$ via the magnetic moment of superconducting rings, we extract $\overline{\rho}_s$ at $T\rightarrow T_c$. The technique, named Stiffnessometer, is sensitive to very small stiffness, which translates to penetration depth on the order of a few millimeters. We apply this method to two different LSCO rings: one with the current running only in the CuO$_2$ planes, and another where the current must cross planes. We find different transition temperatures for the two rings, namely, there is a temperature range with two dimensional stiffness. The Stiffnessometer results are accompanied by Low Energy $\mu$SR measurements on the same sample to determine the stiffness anisotropy at $T < T_c$.

\end{abstract}

\maketitle

%\footnotetext[1]{Correspondence should be addressed to A.K. %(email: keren@physics.technion.ac.il)}

\begin{figure}[tbph]
	\includegraphics[trim=0cm 0cm 0cm 0cm, clip=true,width=\columnwidth]{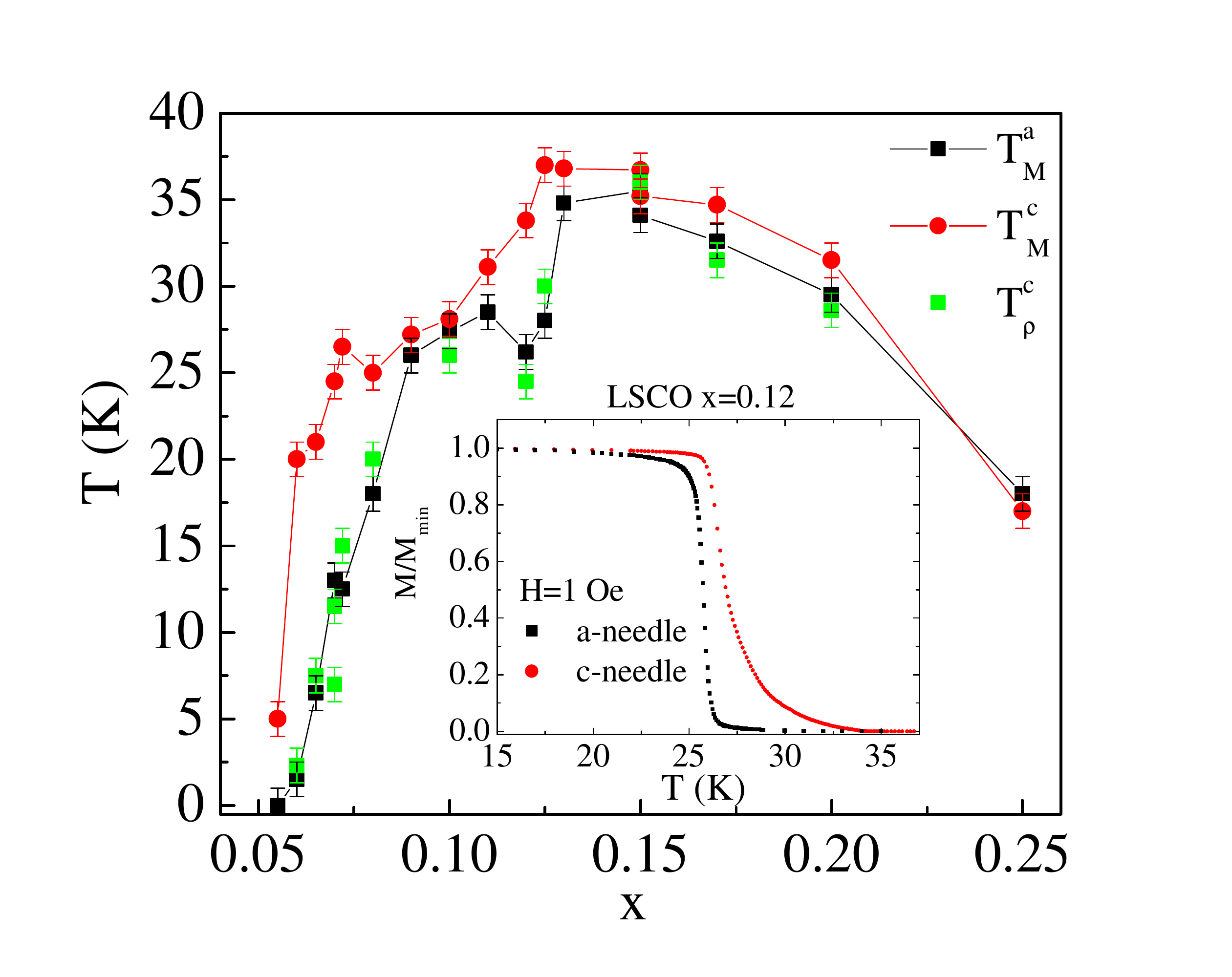}
	\caption{\textbf{LSCO phase diagram.} Temperature versus Sr doping $x$ for a- and c-needles. $T_M$ is the transition temperature taken from magnetization and $T_{\rho}$ is the one taken from resistivity. The inset introduces an example of magnetization measuremen, for two $x=0.12$ needles at $H=1$~Oe. }
	\label{needles}
\end{figure}

The existence of two dimensional (2D) superconductivity (SC) in the CuO$_2$ planes of the cuprates has been demonstrated by either isolated CuO$_2$ sheets~\cite{hetel2007quantum,bovzovic2016dependence}, or in bulk, by applying a magnetic field parallel to these planes~\cite{li2007two,baity2016effective,zhong2018evidence}. In the vicinity of charge stripes formation, the layers are so well decoupled~\cite{berg2007dynamical} that, in fact, two transition temperatures have been found by resistivity~\cite{tee2017two} and magnetization in needle shaped samples \cite{drachuck}, where the demagnetization factor tends to zero, and the measured susceptibility equals the intrinsic one. The magnetization measurements were done in both c-needles, where the CuO$_2$ planes are perpendicular to the field direction, and a-needles where the planes are parallel to the field. An updated phase diagram showing the magnetization critical temperature in c-needles $T_M^c$ and a-needles $T_M^a$ is presented in Fig.~\ref{needles}. The resistivity critical temperature $T_{\rho}^c$ of the same samples agrees with $T_M^a$. The inset shows an example of such magnetization measurement for La$_{2-x}$Sr$_{x}$CuO$_4$ (LSCO) with $x=0.12$.

However, zero resistivity and diamagnetism do not require bulk superconductivity and can occur due to superconducting islands or filaments. It is not clear whether the observed in-plane superconductivity is a macroscopic phenomena and if the sample supports global 2D stiffness as expected from Kosterlitz-Thouless-Berezinski (KTB) theory~\cite{kosterlitz1973ordering,berezinskii1972destruction,kosterlitz1974critical}. If it does, there should be a temperature (and doping) range where the intra-plane stiffness $1/\lambda_{ab}^2$ is finite, while the inter-plane stiffness $1/\lambda_{c}^2$ is zero ($\lambda$ is the penetration depth).

Here we examine the possibility of macroscopic 2D superconductivity in the bulk using two different techniques: Low energy muon spin rotation (LE-$\mu$SR) and Stiffnessometer. The Stiffnessometer is a new method developed to measure particularly small SC stiffness. We focus on the ``anomalous doping" $x=1/8$ regime, where the difference between the two transition temperatures is large, and minute inhomogeneity of Strontium doping does not lead to significant deviations in the transition temperatures.

\begin{figure*}[h!t]
	\includegraphics[trim=0cm 0cm 0cm 0cm, clip=true,width=18cm]{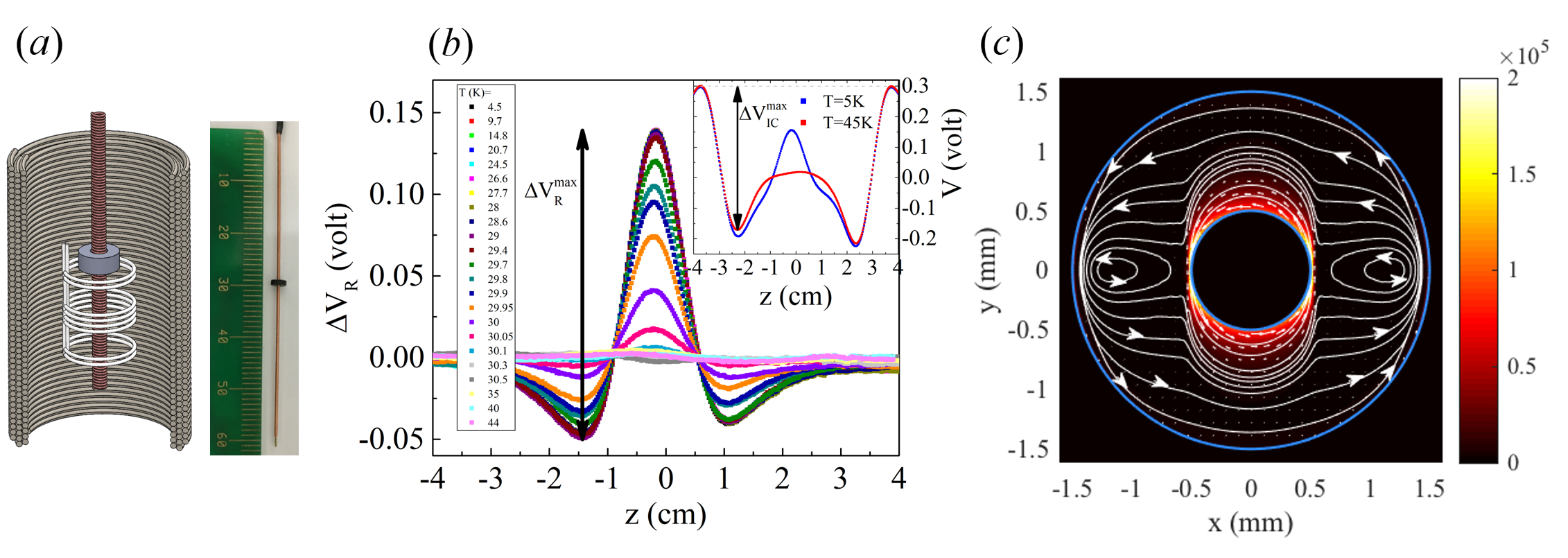}
	\caption{\textbf{Stiffnessometer} (a) An illustration of the Stiffnessometer operation principal and a photo of typical ring and coil with 2400 windings. A long coil is threaded through a ring and they both move with respect to a Gradiometer which is connected to a SQUID. The SQUID measures the flux through the Gradiometer and hence the average vector potential on it $\left\langle A^{\theta}\right\rangle$. (b) Temperature dependence of an LSCO $x=0.125$ c-ring signal as measured by the Stiffnessometer with $I=1$~mA in the inner-coil. The data presented are after subtraction of the coil contribution, $\Delta V_{R}(z)$, as explained in the text and in Ref.~\cite{kapon2017stiffnessometer}. The inset shows raw Stiffnessometer data for a temperature above and below $T_c$. The difference is due to the ring contribution. (c) The currents streamlines in the ring at midheight ($z=0$) derived from the solution of Eq.~\ref{rotrotA} for the a-ring with $\lambda_{c}=145$~$\mu$m and $\lambda_{ab}=13.9$~$\mu$m. The false colors show the current intensity. Naturally the flow is not isotropic. Vortices develop on both sides of the $x$ axis.}
	\label{stiffnessraw}
\end{figure*}

The Stiffnessometer is based on the fact that outside an infinitely long coil, the magnetic field is zero while the vector potential $\textbf{A}$ is finite. When such a coil is threaded through a superconducting ring, the vector potential leads to supercurrent density $\textbf{J}$ according to the London equation ${\bf{J}} = -\overline{\rho}_s  {\bf{A}}$, where $\overline{\rho}_s$ is the stiffness tensor. This current flows around the ring and generates a magnetic moment. We detect this moment by moving the ring and the inner-coil (IC) rigidly relative to a Gradiometer, which is a set of pickup loops wound clockwise and anticlockwise. The Gradiometer is placed in the center of a bigger coil which is used to cancel stray field on the sample. The experimental set-up, our coil and ring are presented in Fig.~\ref{stiffnessraw}(a). The voltage generated in the Gradiometer by the inner coil and the sample movement is measured by a SQUID magnetometer. The measurements are done in zero gauge-field cooling procedure, namely, the ring is cooled to a temperature below $T_c$, and only then the current in the inner coil is turned on. It is the change in magnetic flux inside the inner coil which creates an electric field in the ring, and sets persistent currents in motion.

\begin{figure*}[h!t]
	\includegraphics[trim=5cm 0cm 0cm 0cm, clip=true,width=20cm]{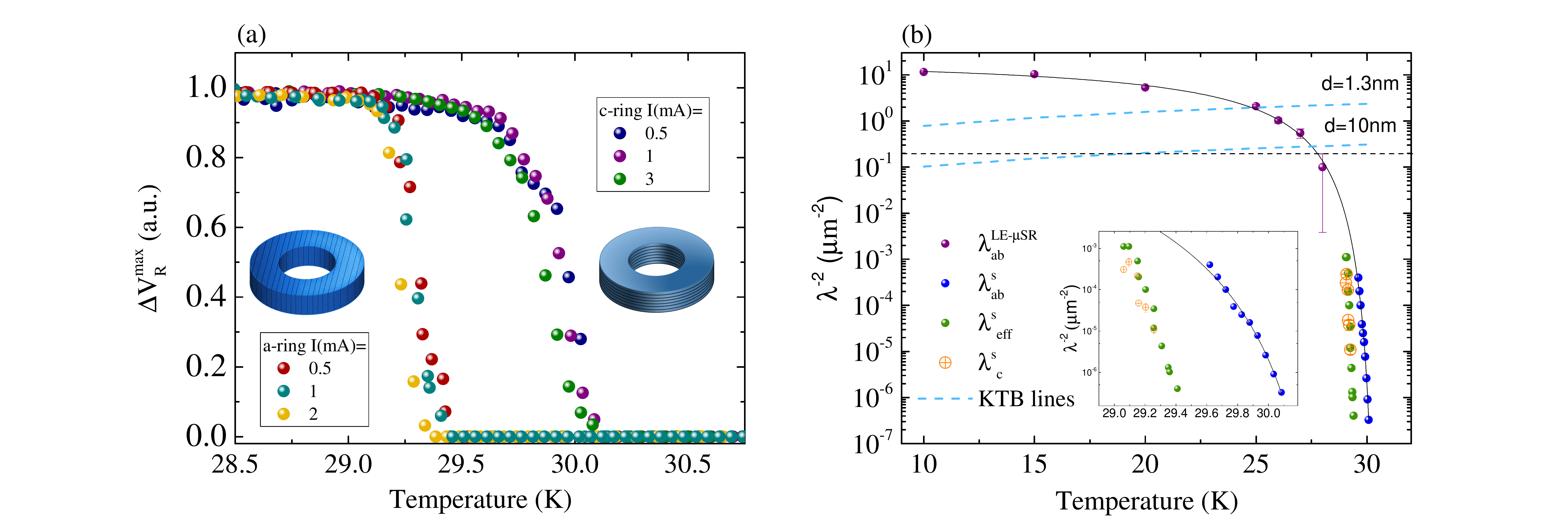}
	\caption{\textbf{LSCO $\mathbf{x=0.125}$ stiffness.} (a) Comparison between a- and c-ring, which are demonstrated in the figure, as measured by the Stiffnessometer. The signal is normalized by the maximum measured ring voltage. Different transition temperatures are observed for the two kind of rings with $0.7$~K difference between them. The transition does not depend on the applied current in the inner coil up to $1$~mA. (b) Semi-log plot of $\lambda_{ab}^{-2}$ as measured by LE-$\mu$SR (purple solid triangles) and Stiffnessometer (blue solid spheres). Black dashed line represents the sensitivity limit of LE-$\mu$SR.  Black solid line is a fit to a phenomenological function described in the text. Dashed blue lines represent the KTB line for layer widths $d=1.3$~nm and $d=10$~nm. Green solid spheres represent the penetration depth of an a-ring from the Stiffnessometer, analyzed as if the ring is isotropic with $\lambda_{eff}$ which is some combination of $\lambda_{ab}$ and $\lambda_{c}$. Orange open symbols show $\lambda_{c}$ obtained at the temperature range where their ratio is manageable numerically for analysis. The inset is a zoom in on temperatures close to the transitions.}
	\label{mainresults}
\end{figure*} 

To examine the orientation dependent response of LSCO to different directions of $\textbf{A}$, we cut two types of rings from a single crystal rod: ``c-ring" where the crystallographic $\hat{\textbf{c}}$ direction is parallel to the ring symmetry axis, i.e. the supercurrent flows in the CuO$_2$ planes, and ``a-ring" where the crystallographic $\hat{\textbf{a}}$ direction is parallel to the ring symmetry axis, i.e. the supercurrent travels both in the planes and between them. The rings, shown in Fig.~\ref{mainresults}(a), have inner radius of $0.5$~mm, outer radius of $1.5$~mm and $1$~mm height.

The inset of Fig.~\ref{stiffnessraw}(b) presents raw Stiffnessometer data of c-ring taken with inner coil current of $1$~mA. The vertical axis is the measured voltage by the SQUID. The horizontal axis is the position $z$ of the ring relative to the center of the Gradiometer. The red data points are measured above $T_{c}$ and represent the signal generated by the inner coil alone. The blue points are measured below $T_{c}$ and correspond to the inner coil and the ring. The difference between them, $\Delta V_R(z)$, is the signal from the ring itself. This signal is shown in Fig.~\ref{stiffnessraw}(b) for different temperatures. Between 4.5~K and 27~K there is hardly any change in the signal, because the Stiffnessometer is not sensitive to short penetration depth compared to the sample size. However, above 28~K the signal drops dramatically fast with increasing temperature.

We define the peak-to-peak voltage of the rings and the inner coil, $\Delta V_{R}^{max}$ and $\Delta V_{IC}^{max}$ respectively, as shown in Fig.~\ref{stiffnessraw}(b). Their ratio holds the information about the stiffness, as we explain shortly. Figure~\ref{mainresults}(a) presents $\Delta V_R^{max}$ of both rings. These voltages are normalized by their maximal value for comparison purposes. We detect two different stiffness transition temperatures, $T_{s}^{c}=30.1$~K for the c-ring, and a lower one $T_{s}^{a}=29.4$~K for the a-ring. We also examine the influence of the inner coil current on the transition. Data corresponding to three different currents are shown in the figure. Below $1$~mA there is no change in the transition, which otherwise widens and appears at slightly lower temperature.

The Stiffnessometer data reveal a new phenomenon. There is a temperature range with finite 2D stiffness in the planes, although supercurrent cannot flow between them. In other words, upon cooling, the SC phase transition starts by establishing a global 2D stiffness, and only at lower temperature a true 3D superconductivity is formed.

To analyze the data, we relate the measured voltage to the vector potential. Since SQUID measures flux, and the vector potential on the Gradiometer is proportional to the flux threading it, the ratio of the peak-to-peak voltages satisfies

\begin{equation}
\frac{\Delta V_R^{max}}{\Delta V_{IC}^{max}}=G\frac{\left\langle A_R^{\theta}(R_{PL} )\right\rangle}{A_{IC}^{\theta}(R_{PL})}
\label{SingleLoop}
\end{equation}
where $A_{R}^{\theta} $ and $A_{IC}^{\theta}$ are the rings and inner coil vector potential components in the azimuthal direction $\mathbf{\hat{\theta}}$ respectively, $R_{PL}$ is the Gradiometer radius, $\left\langle \right\rangle $ stands for averaging over the pickup loops, and $G$ is a geometrical factor determined experimentally (see Supplementary Materials).

In order to extract $\overline{\rho}_{s}$ from the voltages ratio of Eq.~\ref{SingleLoop} we must determine the dependence of ${\bf A}_R(R_{PL})$ on the stiffness. This is done by solving the combined Maxwell's and London's equation

\begin{equation}
\nabla\times\nabla\times\textbf{A}_R = \overline{\rho}_s\left(\textbf{A}_R + \frac{\Phi_{IC}}{2\pi r}\boldsymbol{\hat{\theta}} \right)
\label{rotrotA}
\end{equation}
where $\Phi _{IC}$ is the flux through the inner-coil, and $\overline{\rho}_s$ is finite only inside the ring. For c-ring $\overline{\rho}_s$ is merely a scalar and equals $\lambda_{ab}^{-2}$. For a-ring, it is diagonal in Cartesian coordinates, with $\rho_{xx}=\lambda_{c}^{-2}$ and $\rho_{yy}=\rho_{zz}=\lambda_{ab}^{-2}$.

We solve Eq.~\ref{rotrotA} numerically for our rings geometry and various $\lambda_{ab}$ and $\lambda_{c}$ with FreeFEM++~\cite{freefem} and Comsol 5.3a. The c-ring solution, which is sensitive to $\lambda_{ab}$ only, is discussed in Ref.~\cite{kapon2017stiffnessometer}. Using Eq.~\ref{SingleLoop}, the numerical solution, and the data in Fig.~\ref{stiffnessraw}(b) we extract $1/\lambda_{ab}^2$, and plot it in Fig.~\ref{mainresults}(b) on a semi-log scale (blue solid spheres). 

In order to extract $\lambda_{c}$ we have to know $\lambda_{ab}$ at the temperatures of interest. As can be seen from Fig.~\ref{mainresults}(a) the c-ring Stiffnessometer measurements are in saturation just when a-ring stiffness becomes relevant. Therefore, we applied LE-$\mu$SR to the same samples. 

\begin{figure}[tbph]
	\includegraphics[trim=0cm 0cm 0cm 0cm, clip=true,width=\columnwidth]{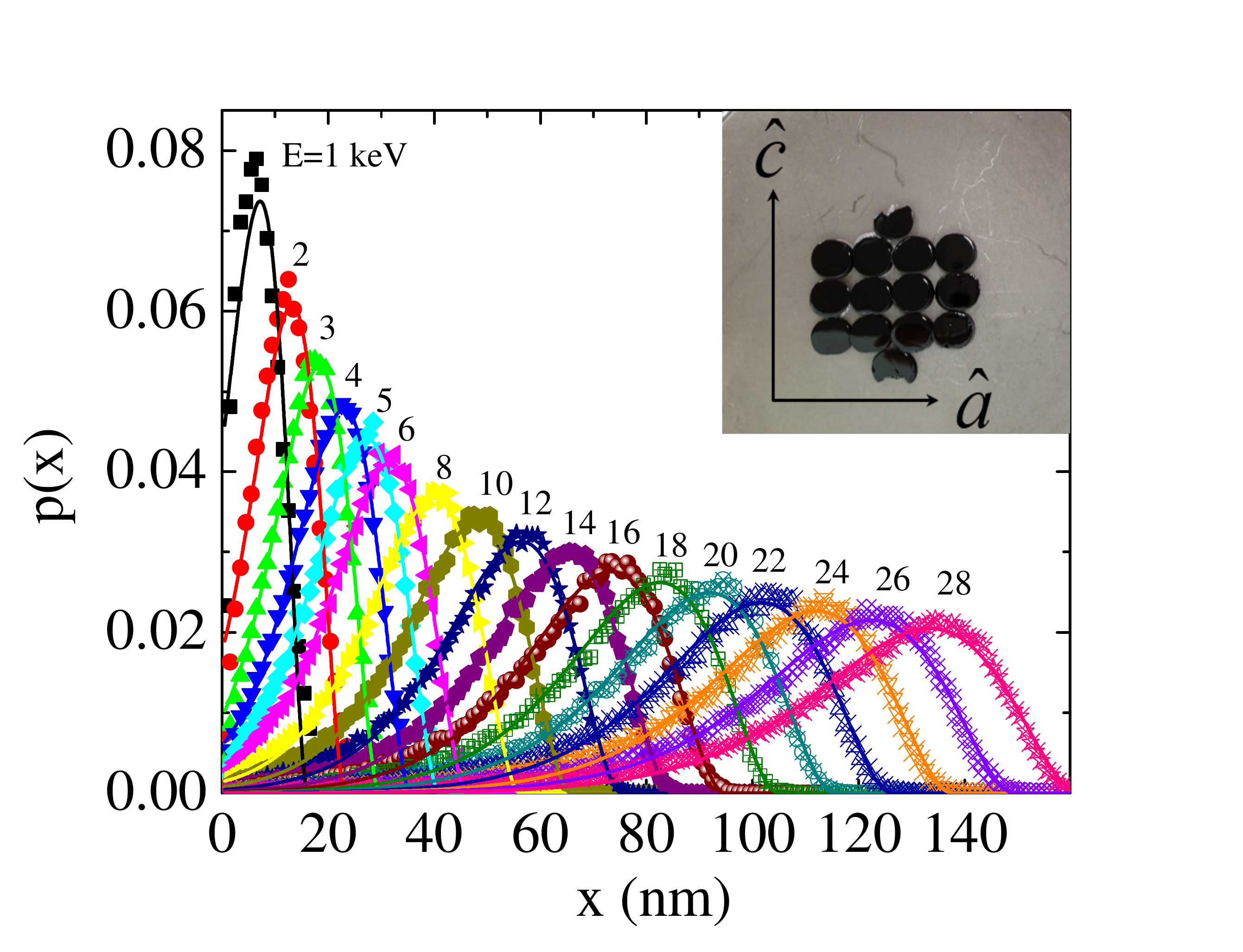}
	\caption{\textbf{Muon stopping profiles.} The probability distribution of a muon to stop at some depth $x$ inside the sample for different implantation energies. The inset shows the LSCO $x=0.125$ single crystal samples used in the experiment. All the pieces were polished to roughness of several nanometers. The crystallographic axes $a$ and $c$ are in the plane of the samples, and shown in the picture. }
	\label{stoppingprofile}
\end{figure}

In LE-$\mu$SR spin polarized muons are injected into a sample. By controlling the muons energy $E$ between 3 to 25 keV, the muons stop with high probability at some chosen depth inside the sample while keeping their polarization intact. The stopping profile $p(x,E)$, where $x$ is stopping depth, is simulated by the TRIM.SP Monte Carlo code~\cite{eckstein2013computer}. Figure~\ref{stoppingprofile} presents the LSCO stopping profiles for different implantation energies. For each energy, we fit the function   

\begin{equation}
\small
p(x,E)=\frac{p_0(x_{0}-x)^3}{\exp[(x_{0}-x)/\xi]-1}H(x_{0}-x).
\label{stopprofile}
\end{equation}  
to this profile. Here $x_{0}$ is some cut-off position the muon cannot cross and is energy dependent, $H(x_{0}-x)$ is Heaviside's function, and $\xi$ and $p_0$ are energy dependent free parameters. The energy dependence of the fit parameters  is given in the Supplemtary Materials section.

When an external magnetic field is applied, the muon spin rotates at the Larmor frequency corresponding to the field. Since the magnetic field decays in the sample on a length scale determined by $\lambda$, the frequency becomes smaller as the muons stop deeper in the sample. We assume an exponential decay of the magnetic field along the direction perpendicular to the sample surface, $x$, resulting from the Meissner effect. In this case, the asymmetry is given by 

\begin{equation}
A(E,t)=A_{0}e^{-t/u}\int_{0}^{\infty}p(x,E)\cos\left(\gamma B_0 e^{-x/\lambda}t\right)dx,
\label{MuonFit}
\end{equation}
where $1/u$ represents contributions to the relaxation from depth independent processes (see Supplementary) and $B_0$ is the mangetic induction outside of the sample and parallel to its surface. For our LE-$\mu$SR measurements, the sample is a mosaic of plates cut in the ac crystallographic plane from the same LSCO $x=0.12$5 crystal used for the Stiffnessometer. Each plate was mechanically polished to a roughness of few tens of nanometers. The plates were glued to a Nickel coated plate using silver paste (see Fig.~\ref{stoppingprofile} inset). We cooled the sample to $5$~K in zero magnetic field. Then a transverse magnetic field was applied along the $\hat{\textbf{a}}$ or $\hat{\textbf{c}}$ directions, and we warmed to the desired temperature. 

\begin{figure}[tbph]
	\includegraphics[trim=0cm 0cm 0cm 0cm, clip=true,width=\columnwidth]{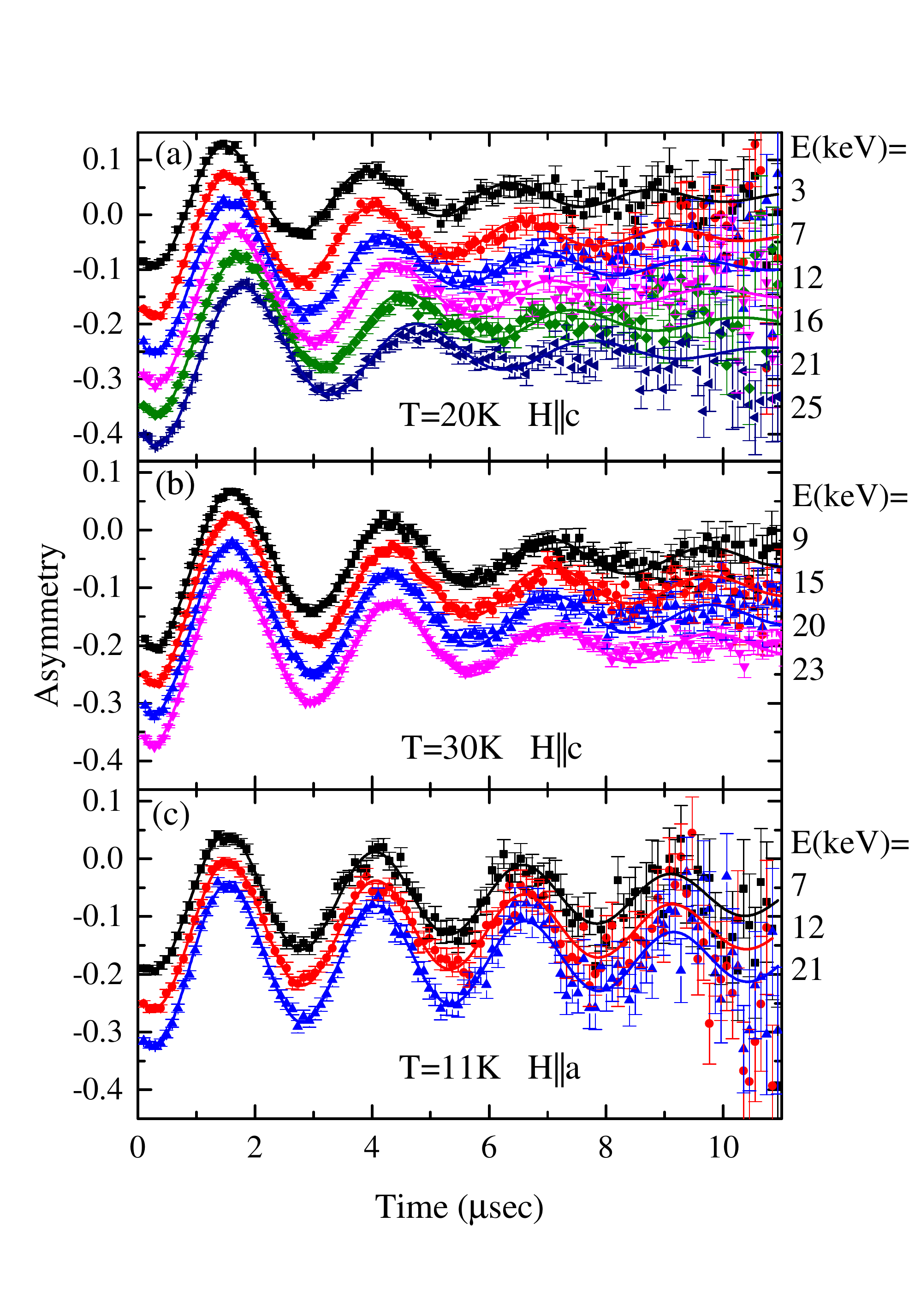}
	\caption{\textbf{LE-$\mu$SR spectra.} Asymmetry as a function of time for different muon implantation energies for: (a) $\textbf{H}\parallel\hat{\textbf{c}}$, $H=26.7$~Oe, $T=20$~K, (b) $\textbf{H}\parallel\hat{\textbf{c}}$, $H=26.7$~Oe, $T=30$~K, (c) $\textbf{H}\parallel\hat{\textbf{a}}$, $H=26.3$~Oe, $T=11$~K. A clear frequency shift as a function of implantation energy is observed in (a). In the (b) conditions, the Stiffnessometer clearly detects stiffness in the ab plane [Fig.~\ref{mainresults}(a)], while no frequency shift is observed by LE-$\mu$SR within our sensitivity. For $\textbf{H}\parallel\hat{\textbf{a}}$ (c) there is no frequency shift at all temperatures.}
	\label{asy}
\end{figure}

Figure~\ref{asy} presents asymmetry data for both magnetic field orientations and different implantation energies. Panels (a) and (b) show data for $\textbf{H}\parallel \hat{\textbf{c}}$ at two different temperatures, and panel (c) depicts data for $\textbf{H}\parallel \hat{\textbf{a}}$. The data sets are shifted vertically for clarity. We limit the presentation to temperatures above $10$~K, since below it strong relaxation due to spin density wave order obscures the oscillatory signal. At $T=20$~K $\textbf{H}\parallel \hat{\textbf{c}}$, we observe a clear frequency shift as a function of implantation energy, indicating a Meissner state. However, for $T=30$~K, where the Stiffnessometer clearly shows $\rho_{ab} > 0$, we could not detect any change in frequency, even though we used high statistics data acquisition of 24 million muons for $E=23$keV and 8 million for the rest. This can be explained from the fact that the penetration depth here is much longer than the muon stopping length scale of the order of hundred nanometer. When $\textbf{H}\parallel \hat{\textbf{a}} $ we did not observe any frequency shift at all temperatures, even though the sample is in the Meissner state.

We fit Eq.~\ref{MuonFit} to our LE-$\mu$SR data and extract $\lambda_{ab}$. We add the results to Fig.~\ref{mainresults}(b). There is a gap between the available data from the two techniques because the longest penetration depth that LE-$\mu$SR can measure, represented by the horizontal dashed line in the figure, is much smaller than the shortest $\lambda$ for which the Stiffnessometer is sensitive to. The function $\lambda_{ab}^{-2}=C_0 \exp \left[C_1/\left(1+C_3 \left(1-T/T_c\right)^\delta\right)\right]$ is fitted to the combined data and serves for interpolation. Since at $T=10$~K we could only measure $\lambda_{ab}$ and not $\lambda_{c}$, we deduce an anisotropy $\lambda_c(0)/\lambda_{ab}(0)\ge10$, as was observed in $\mu$SR, optical, and surface impedance measurements \cite{homes2004universal,shibauchi1994anisotropic,dordevic2005josephson}.

We are now in position to extract $\lambda_{c}$ from Eq.~\ref{SingleLoop}, Eq.~\ref{rotrotA} and the Stiffnessometer a-ring data in Fig.~\ref{mainresults}(a). In this case, two coupled partial differential equations must be solved, where $\lambda_{ab}$ is determined from the c-ring interpolation. Currently, we manage to extract $\lambda_{c}$ for only few temperatures close to $T_s^a$, where the anisotropy ratio is not too big and numerically solvable. These values of $\lambda_{c}$ are presented as orange open symbols in Fig.~\ref{mainresults}(b). The SC currents in the ring at z=0 emerging from the numerical solution for $T=29.16$~K are depicted in Fig.~\ref{stiffnessraw}(c) by combined contour and quiver plots.

For all a-ring Stiffnessometer data we also applied the c-ring stiffness extraction method ignoring the anisotropy. By doing so we determine an effective stiffness $\lambda_{eff}^{-2}$, which is some combination of $\lambda_{ab}^{-2}$ and $\lambda_{c}^{-2}$. These values are presented as green solid spheres in Fig.~\ref{mainresults}(b). $\lambda_{eff}^{-2}$ is larger than $\lambda_{c}^{-2}$ but shows the same trend and indicates two transition temperatures. 

The observation of two transition temperatures is awkward; a material should have only one SC critical temperature. One possible speculation for this result is a finite size effect, namely, if the rings could be made bigger the difference between the two transition temperatures would diminish. This, however, cannot be the case since the sample size is taken into account when extracting the stiffness. Bigger samples should lead to the same $\lambda$ values. A more plausible explanation is that the phase transition starts in the form of wide superconducting filaments~\cite{davis2018spatially} or finite width sheets~\cite{pekker2010finding} in the planes, but disconnected in the third direction. Whether this is the case, or our result indicates a new type of phase transitions, requires further and more local experiments.

The two transition temperatures suggest that there is a temperature range in which the system behaves purely as 2D. Therefore, we examine whether $\lambda_{ab}^{-2}$  follows the KTB behavior. At the KTB transition, the stiffness should undergo a sharp increase (a ``jump") at a temperature $T_{KTB}$ that satisfies $\lambda^{-2}=\gamma T_{KTB}$, where $\gamma = \frac{8k_B e^2 \mu_0}{\pi \hbar^2 d}$ and $d$ is the layer thickness \cite{nelson1977universal}. We plot the line $\lambda^{-2} = \gamma T$ in Fig.~\ref{mainresults}(b) for thickness $d=1.3$~nm of one unit cell (u.c.) and for $d=10$~nm of about 8 u.c., both in cyan dashed lines. Clearly, the KTB line for thickness of one u.c. does not intersect $\lambda_{ab}^{-2}$ where it exhibits a jump. The line for $d=10$~nm, however, does seem to intersect at the beginning of a jump. Thus, for the transition to be of the KTB nature, an effective layer of about $8$ unit cells and more is needed.

In summary, using new magnetic-field-free superconducting stiffness tensor measurements, which are sensitive to unprecedented long penetration depths, on the order of millimeters, and which are not affected by demagnetization factors or vortices, we shed new light on the SC phase transition in LSCO $x=1/8$. In this compound, there is a temperature interval of $0.7$~K where SC current can flow in the CuO$_2$ planes but not between them. When stiffness develops in both directions, the ratio of penetration depths obeys $\lambda_{c}/\lambda_{ab} \geq 10$.

\section*{Acknowledgments}

The Technion physics team is supported by the Israeli Science Foundation (ISF) and by the Technion RBNI Nevet program. The LE-$\mu$SR work is based on experiments performed on the LEM beam line~\cite{prokscha2008new} at the Swiss Muon Source S$\mu$S, Paul Scherrer Institute, Villigen, Switzerland. We are grateful for helpful discussions with Boris Shapiro, Assa Auerbach, Daniel Podolsky, Andreas Suter, and Ori Scally. 

\section*{Supplementary Information}

\subsection*{Materials}

\begin{figure}[tbph]
	\includegraphics[trim=0cm 0cm 0cm 0cm, clip=true,width=\columnwidth]{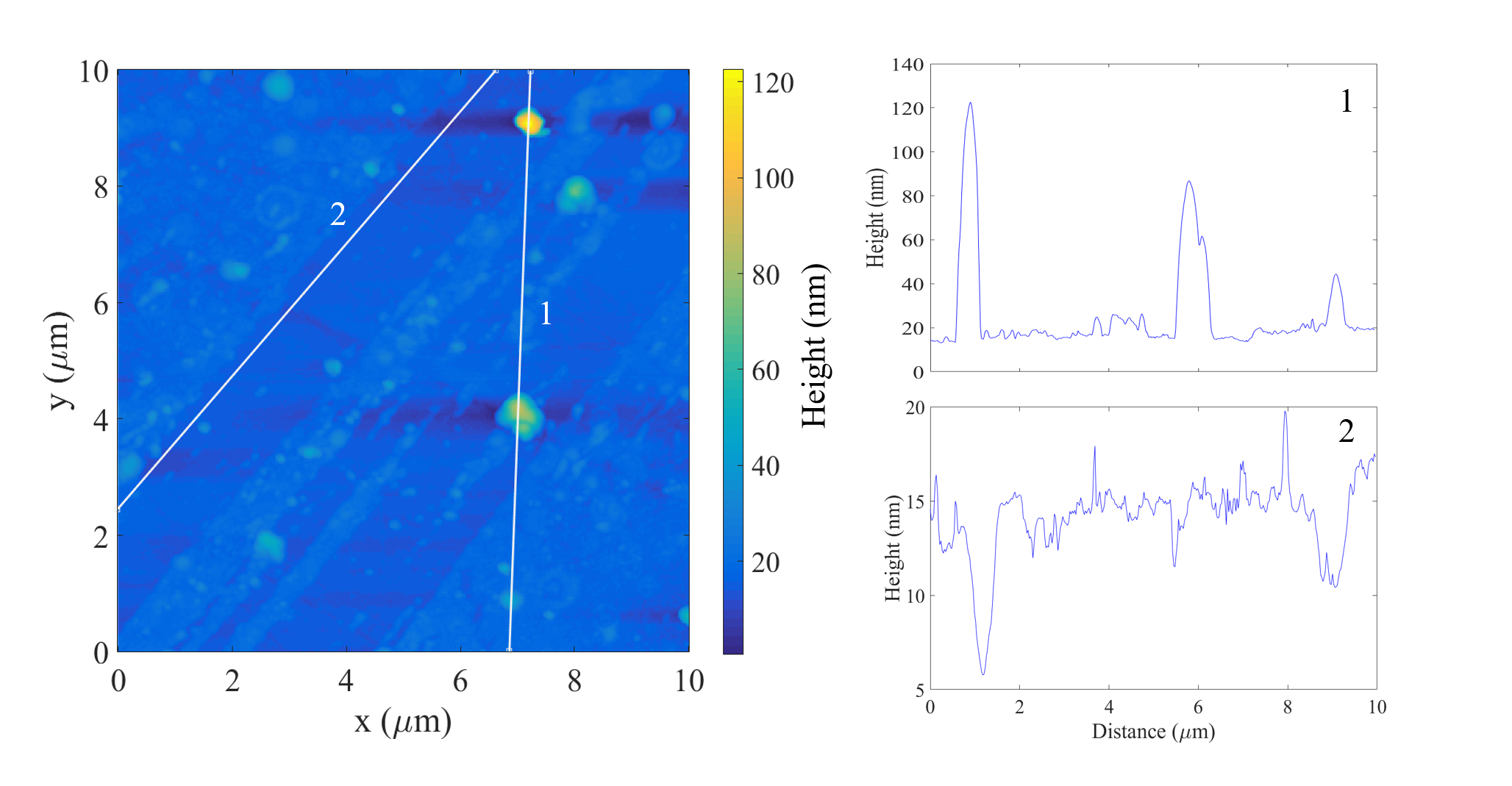}
	\caption{\textbf{Sample surface roughness.} AFM image of one polished LSCO x=0.125 plate treated with $20$~nm Alumina suspension. Height profiles along two lines are presented, demonstrating fairly smooth surface. }
	\label{polish}
\end{figure}

The LSCO single crystals were grown using Traveling Solvent Floating Zone furnace, annealed in Argon environment at $T=850$~C for 120 hours to release internal stress,  and oriented by Laue x-ray diffraction. Stiffnessometer samples were cut into a shape of rings using pulsed Laser ablation, after which the rings were annealed again. LE-$\mu$SR samples were mechanically polished using diamond paste. They were treated eventually with 20 nm alumina suspension. The resulting roughness of few tens of nanometers was determined by Atomic Force Microscope (AFM). A typical AFM data is presented in Fig.~\ref{polish}. 

\subsection*{Stiffnessometer}

The Stiffnessometer is an add-on to a Cryogenic SQUID magnetometer. The components of the experiment shown in Fig.~\ref{stiffnessraw} in the main text are as follows: The inner coil is $60$~mm long with a $0.05$~mm diameter wire and two layers of windings. It is wound on top of a $0.54$~mm diameter polyamide tube. The outer diameter of the coil is $0.74$~mm, and it has 40 turns per millimeter. The second order Gradiometer is $14$~mm high, with inner diameter $25.9$~mm, outer diameter $26.3$~mm, and made from $0.2$~mm diameter wire. We take $R_{PL}=13\pm0.15$~mm. The Gradiometer is constructed from three groups of windings distanced $7$~mm apart from each other. The upper and lower ones have two loops wound clockwise, while the center windings have four loops wound anticlockwise. Numeric evaluation of the G factor in Eq.~\ref{SingleLoop} using the Gradiometer dimensions gives a reasonable result for an isotropic superconducting ring with known dimensions~\cite{kapon2017stiffnessometer}.

For anisotropic ring the situation is much more complicated. Therefore, the G factor used here is extracted experimentally. As shown in Fig.~\ref{mainresults} in the main text, the signal from the rings $\Delta V_{R}^{max}$ saturates at $T \ll T_c$. It happens when the penetration depth is much smaller than the ring dimensions. The ratio between the voltages saturation value to the vector potentials ratio calculated numerically  gives G.

\begin{figure}[tbph]
	\includegraphics[trim=0cm 0cm 0cm 0cm, clip=true,width=\columnwidth]{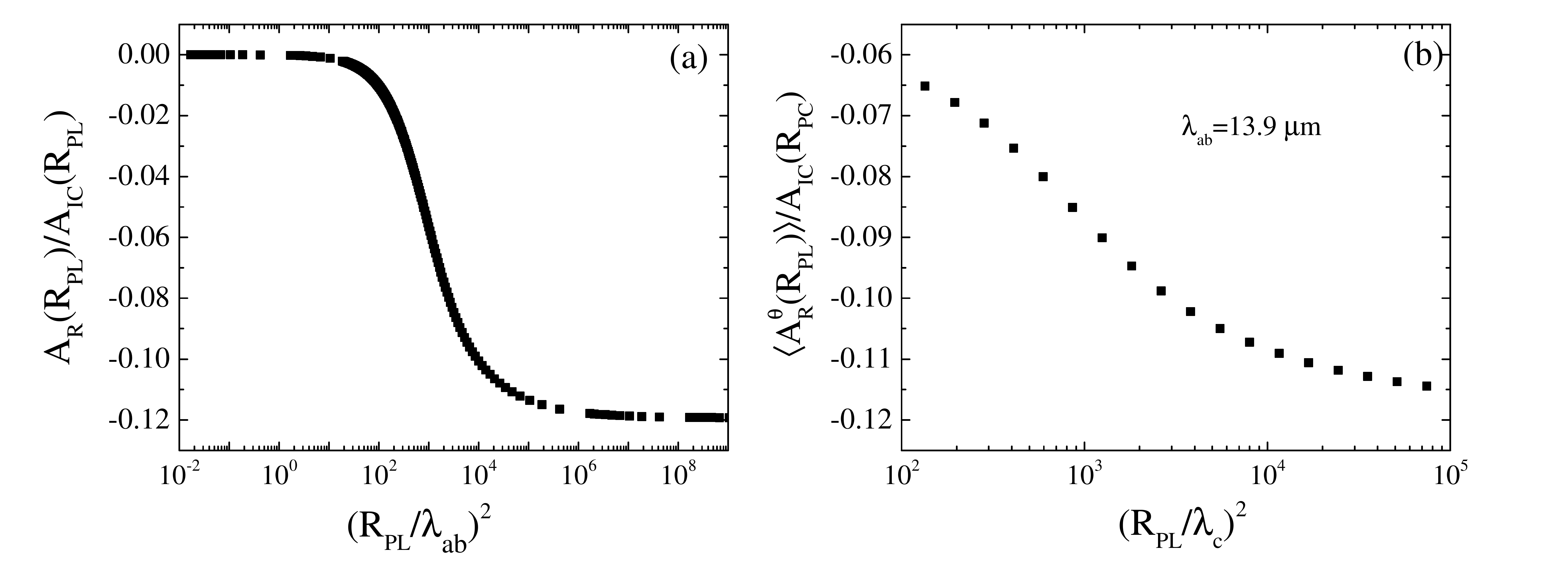}
	\caption{\textbf{Extracting the stiffness.} Numerical results of the vector potentials ratio as appears in Eq.~\ref{SingleLoop} as a function of (a) $(R/\lambda_{ab})^2$ and (b) $(R/\lambda_{c})^2$ for $\lambda_{ab}=13.9$~$\mu m$ at $T=29.16$~K.}
	\label{Anumerical}
\end{figure} 

\subsection*{Numerical methods}

\begin{figure*}[tbph]
	\includegraphics[trim=0cm 0cm 0cm 0cm, clip=true,width=18cm]{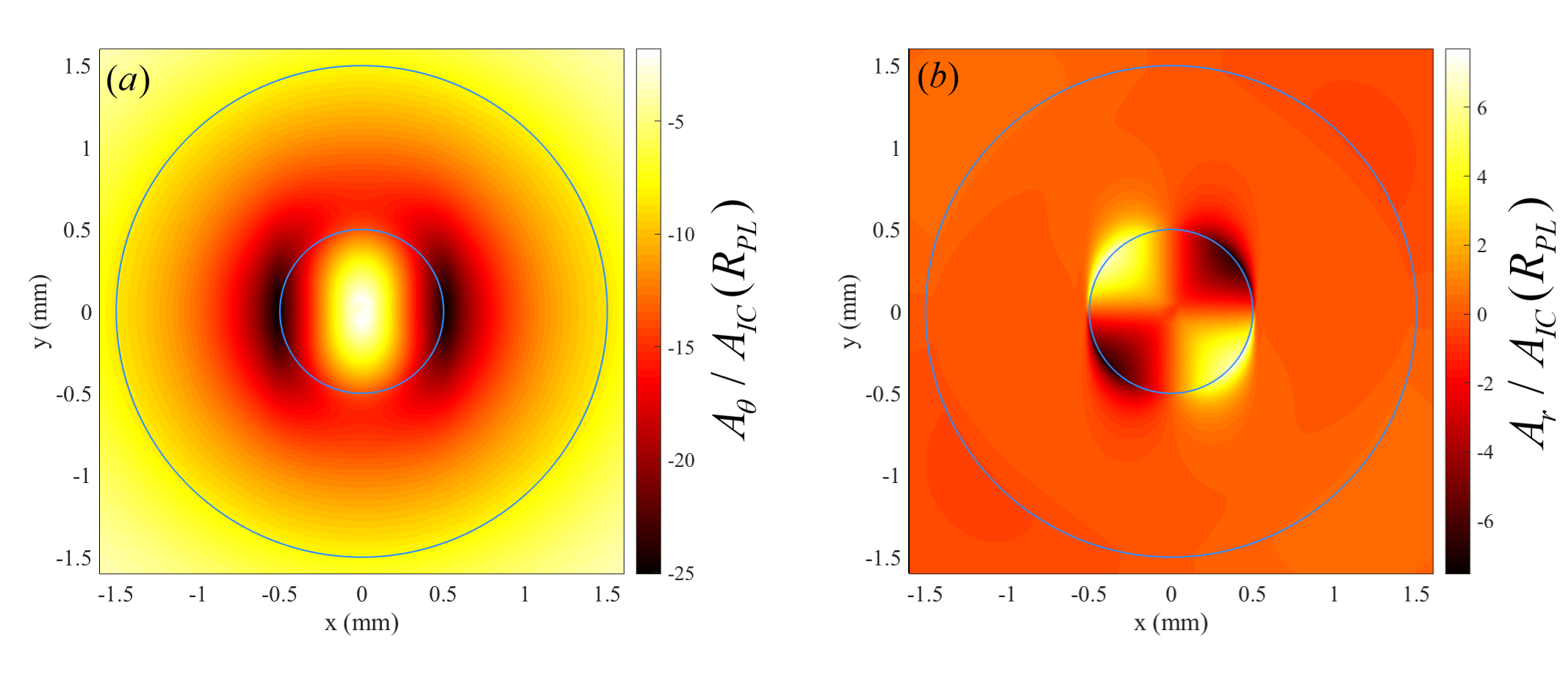}
	\caption{\textbf{Vector potential for LSCO x=0.125 a-ring.} Numeric solution of the radial (a) and azimuthal (b) components of the vector potential inside the ring at z=0 for $\lambda_{ab}=13.9$~$\mu m$ and $\lambda_{c}=145$~$ \mu m$}
	\label{Aanisotropic}
\end{figure*} 

Here we provide more details about the numeric solution of Eq.~\ref{rotrotA} in the anisotropic case. The gauge choices are as follows: Inside the ring, applying divergence to Eq.~\ref{rotrotA} yields the gauge $\nabla\cdot\left(\overline{\rho}_s\mathbf{A}_{tot}\right) = 0$, where $\mathbf{A}_{tot}=\mathbf{A_R} + \mathbf{A_{IC}}$. This gauge also enforces the continuity equation for the current density $\mathbf{J}=\overline{\rho}_s\mathbf{A}_{tot}$. Outside the ring we apply the Coulomb gauge $\nabla \cdot\mathbf{A}_{tot} = 0$, which is also used to determine $\mathbf{A_{IC}}$ and $\mathbf{A}_{tot}$ in the isotropic case. The boundary conditions are $A(\infty)=0$. In practice, infinity is understood as the domain surface, and the domain is taken to be large enough so that finite-domain effects are negligible. The domain of the problem is defined as a cylinder with height 100 times that of the ring, i.e. 7.7$R_{PL}$ and outer radius 100 times that of the ring, i.e. 11.5$R_{PL}$. Since no current can cross the ring surface, we demand ${{\mathbf{J}}_ \bot }({{\mathbf{r}}_{in}}) = {{\mathbf{J}}_ \bot }({{\mathbf{r}}_{out}}) = 0$ where $\bot$ stands for the direction perpendicular to the surface, and $r_{in}$ ($r_{out}$) is the inner (outer) radius of the ring. Finally, from the absence of a surface field, we demand $\Delta {{\mathbf{A}}_\parallel }({{\mathbf{r}}_{in}}) = \Delta {{\mathbf{A}}_\parallel }({{\mathbf{r}}_{out}}) = 0$,
where $\Delta {{\mathbf{A}}_\parallel }$ stands for the difference between the vector potential parallel to the surface inside the ring and outside of it.

%Combining all these together with Eq. ??? from the main text generates the following two coupled equations inside the ringe: 
%\begin{widetext}
%\[\begin{gathered}
%{\nabla ^2}{A_r} - \frac{{{A_r}}}{{{r^2}}} - \frac{2}{{{r^2}}}\frac{{\partial {A_\theta }}}{{\partial \theta }} = {A_r}\left[ {{\rho _{xx}}{{\cos }^2}\theta  + {\rho _{yy}}{{\sin }^2}\theta } \right] + \left[ {\frac{{{\rho _{yy}} - {\rho _{xx}}}}{2}} \right]\left[ {{A_\theta } + \frac{1}{r}} \right]\sin 2\theta  \hfill \\
%{\nabla ^2}{A_\theta } - \frac{{{A_\theta }}}{{{r^2}}} + \frac{2}{{{r^2}}}\frac{{\partial {A_r}}}{{\partial \theta }} = {A_r}\left[ {\frac{{{\rho _{yy}} - {\rho _{xx}}}}{2}} \right]\sin 2\theta  + \left[ {{\rho _{xx}}{{\sin }^2}\theta  + {\rho _{yy}}{{\cos }^2}\theta } \right]\left[ {{A_\theta } + \frac{1}{r}} \right] \hfill \\ 
%\end{gathered} \]
%\end{widetext}
%
%whereas outside of it $\nabla^2 \mathbf{A} = 0$.

Figure~\ref{Anumerical} shows the numerical results of the vector potentials ratio that appears in Eq.~\ref{SingleLoop} as a function of (a) $(R/\lambda_{ab})^2$ and (b) $(R/\lambda_{c})^2$ for $\lambda_{ab}=13.9$~$\mu m$ at $T=29.16$~K.  In our analysis, $\lambda_{ab}$ is extracted from the c-ring data in the isotropic case. Then, for each temperature, the corresponding $\lambda_{ab}$ is used to generate the result in panel (b), and combining with the a-ring data $\lambda_{c}$ is extracted.

Figure~\ref{Aanisotropic} presents the numeric solution of the ring vector potential $A_R$ at $z=0$ plane (midheight of the ring), calculated for LSCO x=0.125 a-ring at $T=29.16$~K with $\lambda_{ab}=13.9$~$\mu m$, extracted from the extrapolation function presented in the main text, and $\lambda_{c}=145$~$\mu m$. Panel (a) shows the azimuthal part of $\mathbf{A}$, whereas panel (b) the radial one.

\begin{figure*}[tbph]
	\includegraphics[trim=0cm 0cm 0cm 0cm, clip=true,width=18cm]{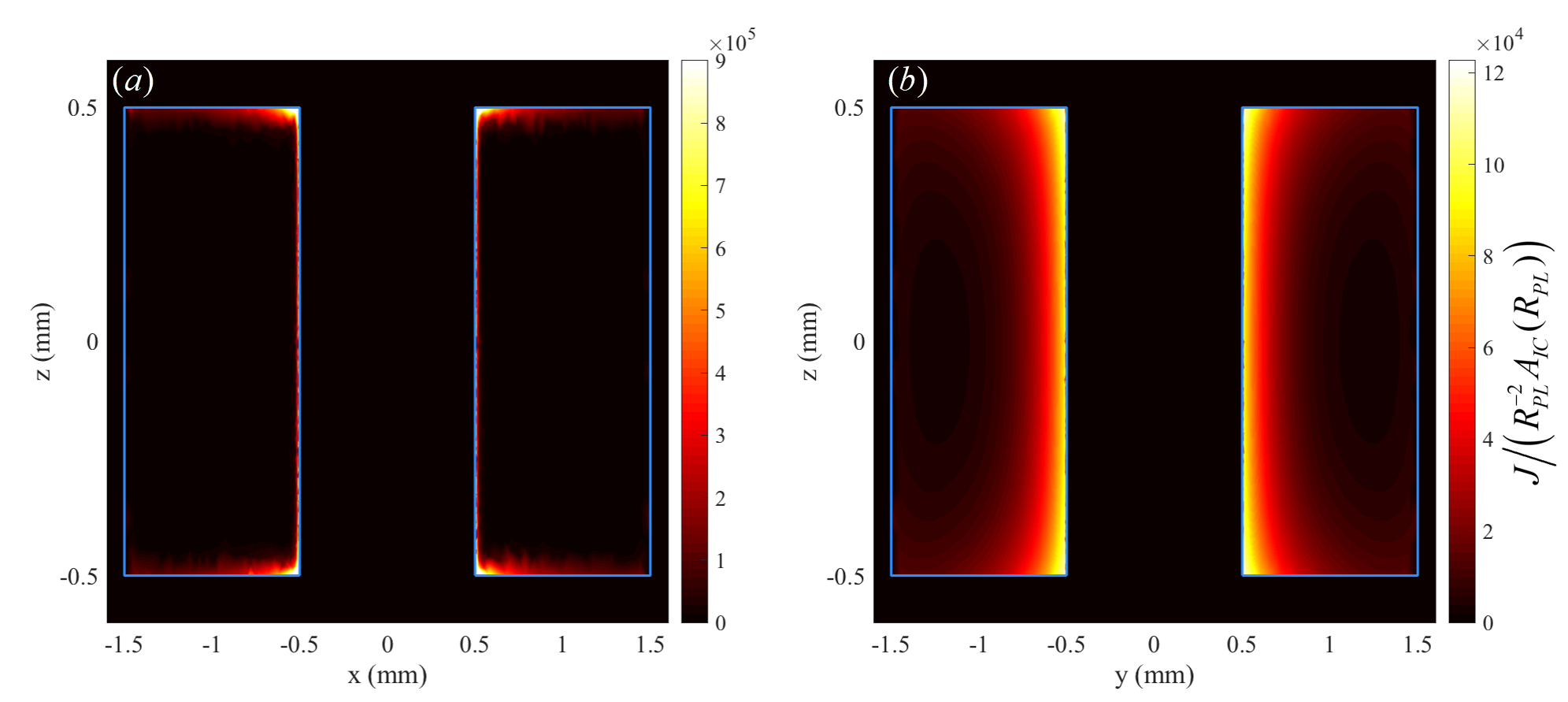}
	\caption{\textbf{Current density simulation inside LSCO x=0.125 a-ring.} False color map of the current density distribution in a ring with $\lambda_{ab}=13.9$~$\mu m$ and $\lambda_{c}=145$~$\mu m$ in the (a) xz plane and (b) yz plane. Most of the current concentrates on the inner rim. }
	\label{Janisotropic}
\end{figure*}

Figure~\ref{Janisotropic} shows the absolute value of the current density $\mathbf{J}$ inside the rings for two cuts at fixed angeles: (a) xz plane, (b) yz plane. At the xz plane, the current concentrates at a very thin layer close to the ring inner rim, while in the yz plane the current penetrates further into the bulk. This corresponds, of course, to the large difference in the penetration depth in the two directions. 

\begin{figure}[tbph]
	\includegraphics[trim=0cm 0cm 0cm 0cm, clip=true,width=\columnwidth]{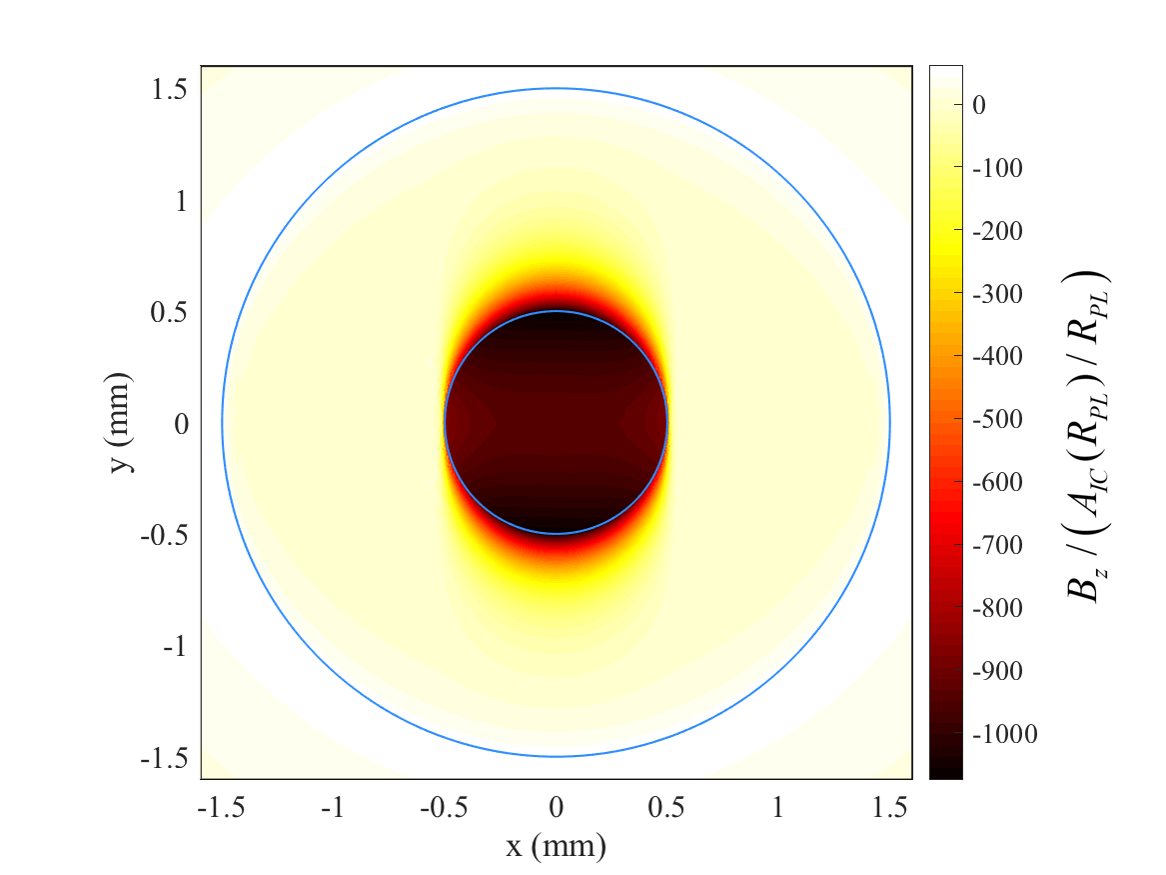}
	\caption{\textbf{Magnetic field} False color map of the magentic field z component inside the a-ring and its vicinity for $\lambda_{ab}=13.9$~$\mu m$ and $\lambda_{c}=145$~$\mu m$ }
	\label{Banisotropic}
\end{figure}

Finally, Fig.~\ref{Banisotropic} shows the magnetic field generated by the ring as calculated from the curl of $\mathbf{A_{R}}$. The penetration pattern of the field is of an ellipse due to the penetration depths anisotropy.

\subsection*{LE-$\mu$SR}
%In a standard $\mu$SR experiment, a polarized beam of muons with $E=4$~MeV ???? is injected into the sample, where the muons spin precess according to the inner microscopic magnetic field inside it. Having average life time of approximately 2.2 $\mu$~s, the muon decays into a positron, neutrino and anti-neutrino. The emitted positrons are detected in two detectors in the sample vicinity, Backward and Forward, and the time dependent polarization of the muon in constructed from the asymmetry between the signals. Given the gyromagnetic ratio of the muon $\gamma=135.5$~MHz/T, the average magnetic induction in the sample is calculated using the asymmetry frequency.

In the LE-$\mu$SR experiment, 4 MeV spin-polarized muons are stopped at a moderator, made of 300 nm thick layer of solid Argon grown on top of a silver foil. They are then accelerated to a chosen energy between 1 to 30 keV by applying a voltage difference between the foil and the sample. The sample holder is placed on a sapphire plate hence electrically isolated. The parameters of the stopping profiles given in the main text by Eq.~\ref{stopprofile} are   
\begin{gather*}
p_0(E) = \exp\left[-6.4-0.8\ln E-0.18(\ln E)^2\right] \\ 
x_{0}(E) = 12+6E-0.11E^2+0.0028E^3 \\
\xi(E) = 2.77+0.49E-0.0165E^2+0.0003E^3 \\
\end{gather*}
The whole chamber is under ultra high vacuum of $10^{-10}$~mbar, and the stopping and accelerating proceses  of the muons preserve most of the polarization. Once in the sample, the muon spin rotates in the local external or internal magnetic field and the time dependent polarization is reconstructed from asymmetry in the positrons decay, which are emitted preferentially in the muon spin direction.

\begin{figure}[tbph]
	\includegraphics[trim=0cm 0cm 0cm 0cm, clip=true,width=\columnwidth]{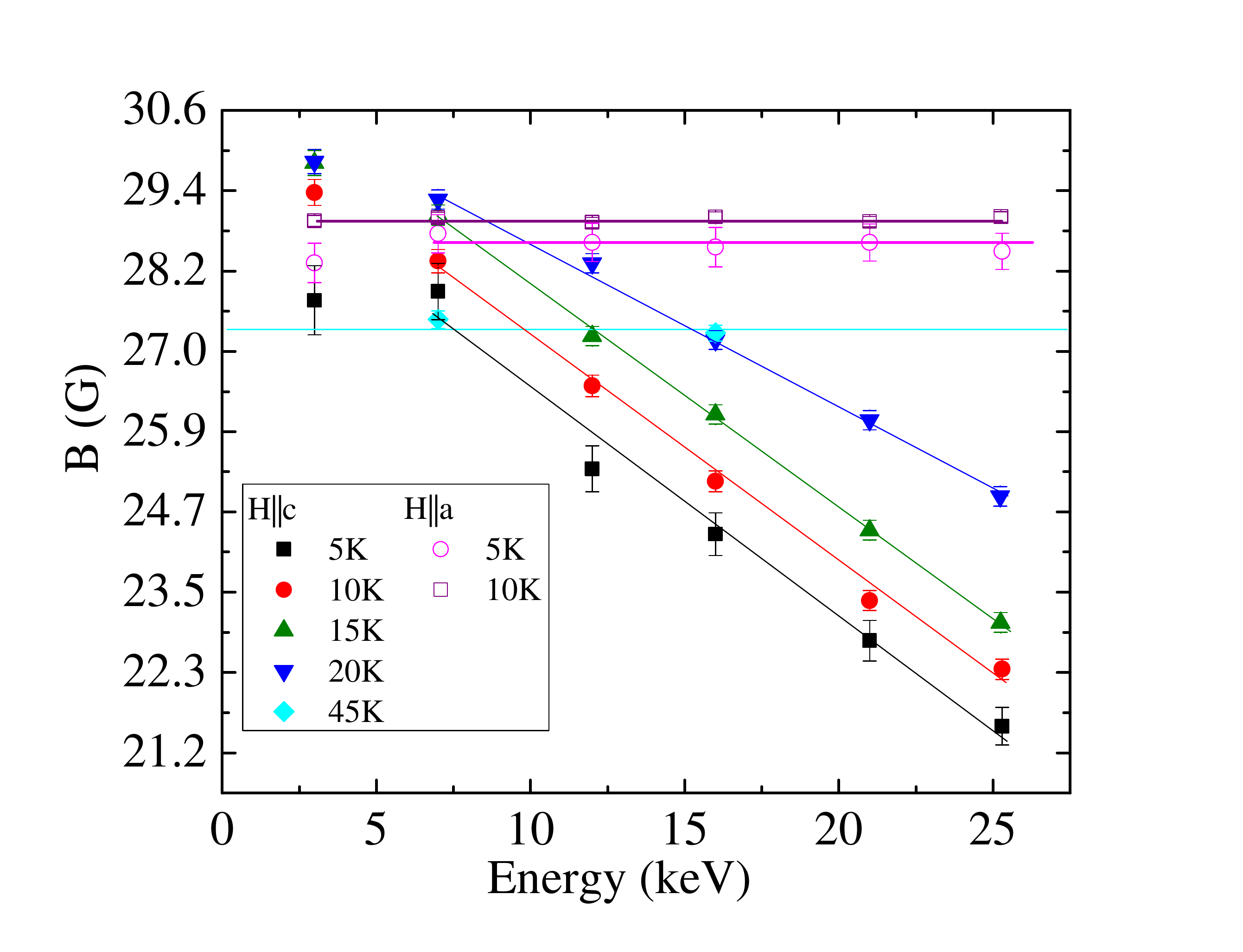}
	\caption{\textbf{Magnetic field as function of implantation energy.} Closed symbols are $\mathbf{H}\parallel \mathbf{c}$ and open symbols are $\mathbf{H}\parallel \mathbf{a}$. Straight lines are guides to the eye. The magnetic field below $E=5$~keV does not fit the linear trend of the field decay, indicating a dead layer of about 10 to 20 nanometers, possibly caused by the polishing treatment.}
	\label{DeadLayer}
\end{figure}

There are two methods by which one can extract the penetration depth. The simple method is to fit each data set (at each temperature and energy) to $A(t) = A_0\exp(-t/T_2) \cos(\omega t)$. From this fit one can extract asymmetry, relaxation, and the average internal field as a function of average implantation depth and temperature. Figure~\ref{DeadLayer} summarizes the internal magnetic field as a function of implantation energy for different temperatures and field orientations. The field here is calculated by $B = \omega / 2\pi\gamma$, where $\omega$ is the angular frequency of the muon polarization and $\gamma$ is the gyromagnetic ratio. Noticeably, close to the surface and at low $T$, the magnetic field does not change with increasing implantation depth for $\mathbf{H} \parallel \mathbf{c}$. Only for energies above $5$~keV does a linear trend of decay appears. This $10$ to $20$ nanometers of ``dead layer" could be a byproduct of the polishing process.

\begin{figure}[tbph]
	\includegraphics[trim=0cm 0cm 0cm 1.5cm, clip=true,width=\columnwidth]{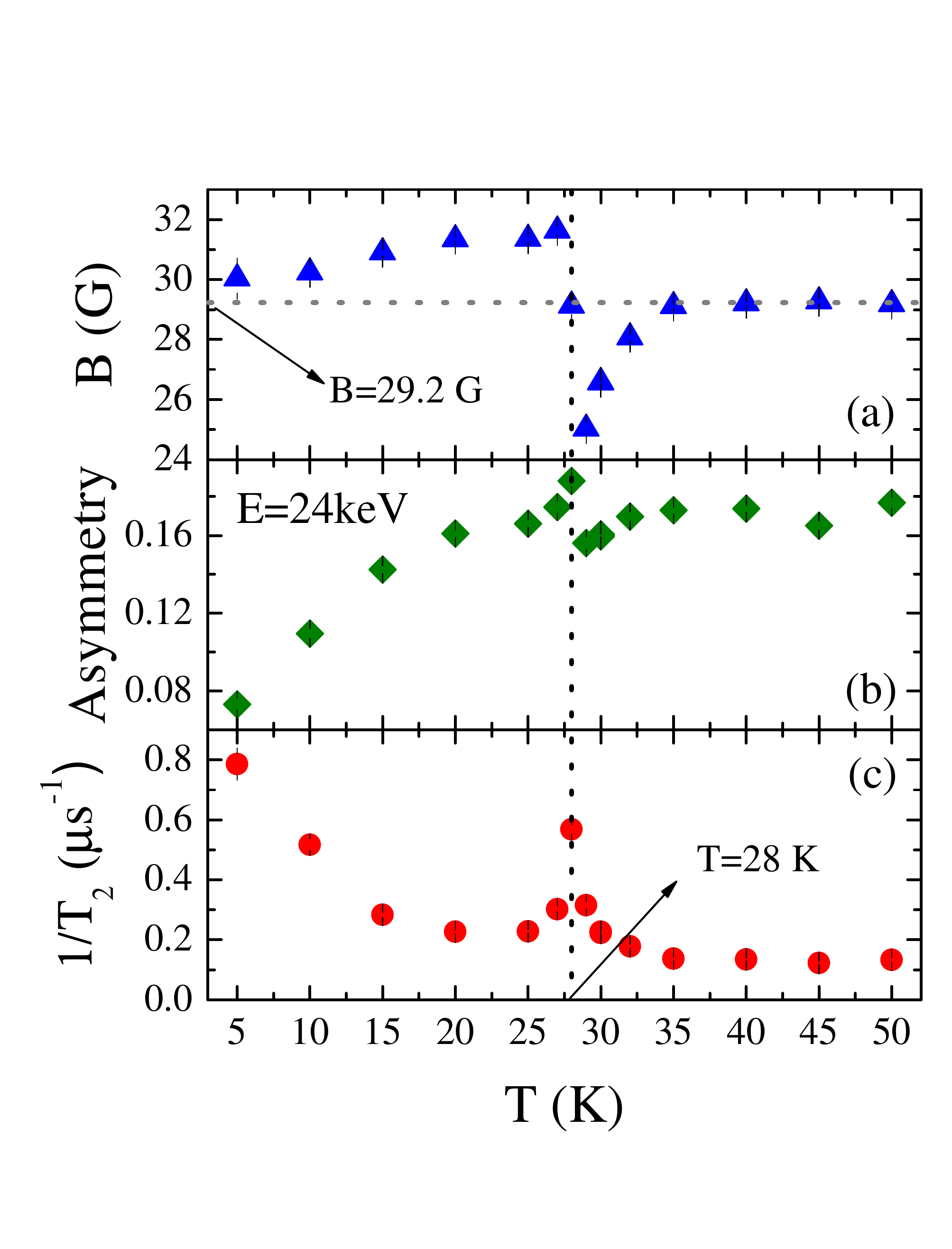}
	\caption{\textbf{Temperature dependance of LE-$\mu$SR parameters}. The measurement was done at constant energy of 24 keV. The magnetic field (a) displays peculiar behavior near $T_c$. Its magnitude below $T_c$ is larger than that of the normal state. The asymmetry (b) is constant until 20K, where it starts to drop due to magnetic freezing. The magnetism is also exhibited in an uprise of the decay rate (c) at low temperatures.}
	\label{LemTdep}
\end{figure}

Figure~\ref{LemTdep} depicts the temperature dependence of the individual fit parameters for the highest implantation energy. The magnetic field (panel (a))seems to behave erratically close to the phase transition into the superconducting state. We attribute this behavior to demagnetization factor and mutual coupling between different pieces of the sample. The asymmetry (panel (b)) decreases upon cooling since LSCO x=0.125 is known to have a magnetic phase concomitant with the superconducting one~\cite{panagopoulos2002evidence,PhysRevB.57.R3229,PhysRevB.59.6517}. The muon spin relaxation (panel (c)) has a peak at the critical temperature, which is also unusual.

\begin{figure}[tbph]
	\includegraphics[trim=0cm 0cm 0cm 0cm, clip=true,width=\columnwidth]{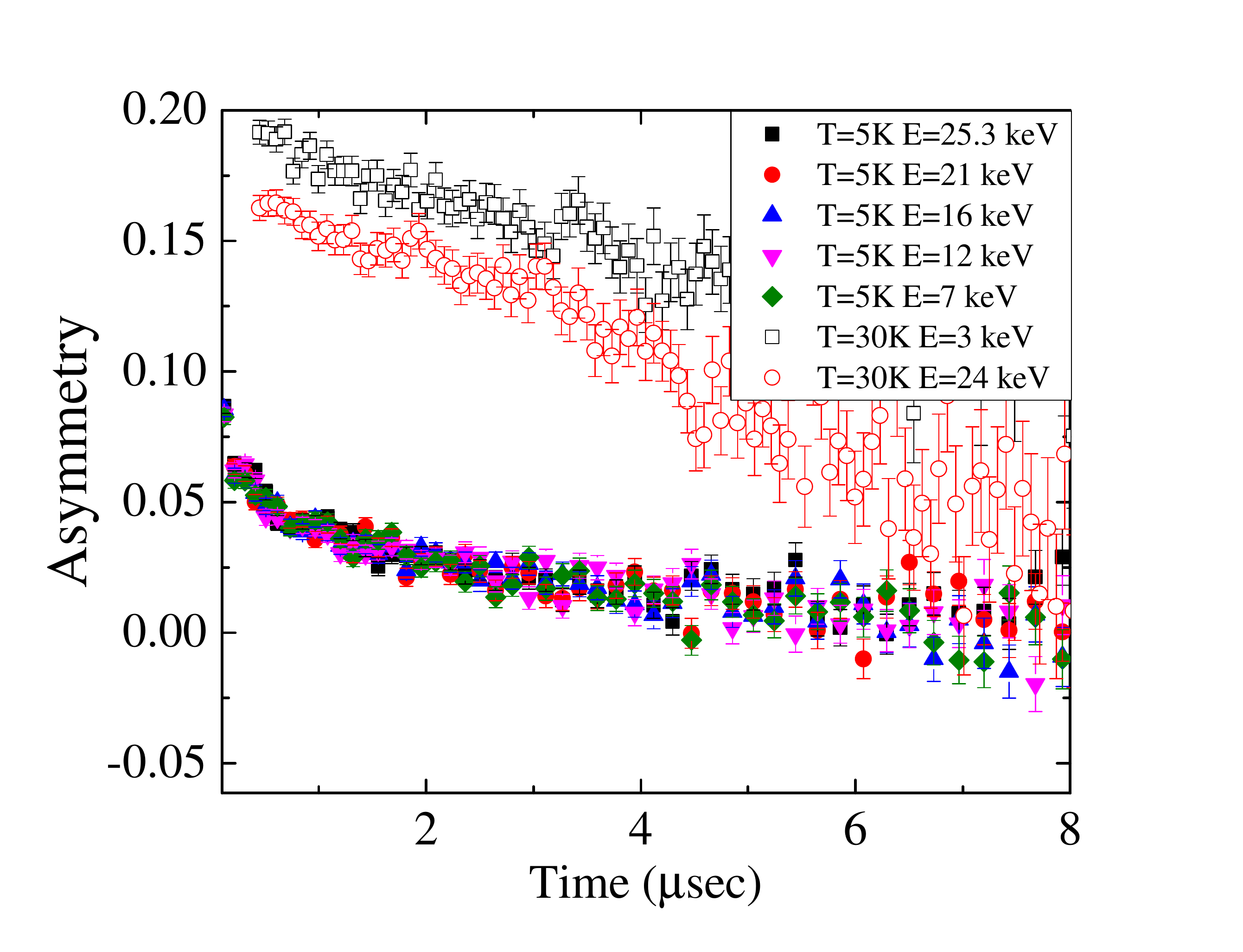}
	\caption{\textbf{Depth independent magnetism in LSCO x=0.125}. Asymmetry vs. time at $T=5$~K (close symbols) and $T=30$~K (open symbols) for different implantation energies. The signal does not change as a function of energy at low temperatures, justifying a depth independent relaxation component (see main text).}
	\label{magnetism}
\end{figure}

The presence of magnetism could be detrimental to our analysis if it depends on depth. To verify that this is not the case, we perform zero field (ZF) measurements for different implantation energies at $T=5$~K well below $T_c$ and for $T=30$~K above $T_c$. The results are presented in Fig.~\ref{magnetism}. Fast relaxation and reduction of the asymmetry are observed at low temperature due to local random fields originating from the magnetic stripes in the sample. Nevertheless, there is no change in the magnetic relaxation with implantation depth.

The more sophisticated analysis method is presented in the main text. For each temperature, we fit all data sets with energy larger than $5$~keV due to the presence of a dead layer, using Eq.~\ref{MuonFit}. In the fit $A_0$ is a free parameter, and $\lambda$, $u$ and $B_0$ are shared. $A_0$ is free because the number of muons actually penetrating the sample varies with energy. $u$ represents relaxation processes that are implantation depth independent such as magnetism or field variations perpendicular to $x$. These are taken into account as some Lorentzian probability distribution of the total internal magnetic field with FWHM of $2/u$. $\lambda$ and $B_0$ are naturally common to each temperature. Comparing the two analysis methods for $T=25$~K, for example, the penetration depths agree within 20$\%$.

%\bibliography{references}
%\bibliographystyle{naturemag_noURL}

\end{document}